\documentclass[aps,prd,twocolumn,showpacs,amsmath,amssymb,amsthm,nofootinbib, preprintnumbers]{revtex4-1}
 \usepackage{amsmath}
\usepackage{graphicx}
\usepackage{epstopdf}
\usepackage{float}
\usepackage{hyperref}
\usepackage{color}
\usepackage[T1]{fontenc}
\usepackage[utf8]{inputenc}
\usepackage[toc,page]{appendix}
\usepackage[usenames,dvipsnames]{xcolor}
\usepackage[normalem]{ulem}
\usepackage{lipsum, babel}
\usepackage{amsthm}

\newcommand{\be}{\begin{equation}}
\newcommand{\ee}{\end{equation}}
\newcommand{\ba}{\begin{eqnarray}}
\newcommand{\ea}{\end{eqnarray}}

\newcommand{\beq}{\begin{equation}}
\newcommand{\eeq}{\end{equation}}
\newcommand{\beqa}{\begin{eqnarray}}
\newcommand{\eeqa}{\end{eqnarray}}

\newtheorem*{theorem}{Theorem}

\begin{document}

\title{{Higher-Dimensional Slowly Rotating Black Holes in Nonlinear Electrodynamics}}

\author{Tayebeh Tahamtan}
\email{tayebeh.tahamtan@matfyz.cuni.cz}

\affiliation{Institute of Theoretical Physics, Faculty of Mathematics and Physics,
Charles University, Prague, V Hole{\v s}ovi{\v c}k{\' a}ch 2, 180 00 Prague 8, Czech Republic}

\date{\today}

\begin{abstract}

We study slowly rotating, nonlinearly charged black holes in arbitrary spacetime dimension $d \geq 4$, using a generalized Lense--Thirring ansatz. Working to linear order in the rotation parameters, we solve the Einstein--nonlinear electrodynamics (NLE) field equations for restricted theories ${\cal L}({\cal S})$ depending on single invariant. We provide an extension of the four-dimensional "No Go theorem" proven in ArXiv:2203.01919  and demonstrate the uniqueness of Maxwell theory in $d \geq 4$.  Similar to four dimensions, it is proven that the standard rotation generating trick fails to produce full rotating solutions for any nontrivial NLE in higher dimensions. We identify a family of NLE theories admitting slowly rotating solutions with vector potentials encoding rotation effect in Maxwell-like manner; in four dimensions this reproduces the previously found RegMax theory, while in $d\geq4$ we obtain the corresponding electric potential in closed form and the associated Lagrangian implicitly as a function of the radial coordinate.

\end{abstract}

\maketitle

\section{Introduction}
{In general relativity, rotating black holes are generally considered to be the best model for sources of many astrophysical phenomena --- gravitational waves, processes in the centers of galaxies, relativistic jets, etc. However, finding such black holes analytically is a challenging task. The first rotating black hole to solve the vacuum Einstein field equations was discovered in 1963 by R. Kerr \cite{Kerr1963} and was later named the Kerr black hole. The natural extension to a charged, rotating black hole - the Kerr--Newman metric- was discovered shortly thereafter in 1965 \cite{Newman:1965tw} using a technique later called the Newman--Janis trick.  These were studied in four-dimensional spacetime. A key question (purely theoretical but also motivated by theories involving extra dimensions) is whether such compact objects exist in higher dimensions, and what their physical properties are. Could a generalization of the Newman–Janis trick be used to find a fully rotating solution in higher dimensions with any type of physical source?

The first higher-dimensional black hole solutions were found by Tangherlini in 1963 \cite{Tangherlini:1963bw}. These spacetimes generalized the spherically symmetric Schwarzschild and Reissner--Nordstr\"{o}m solutions in four dimensions.  The first generalization of rotating black holes in higher dimensions was found by Myers and Perry \cite{MYERS1986}.  Later, in \cite{X-Dianyan1988}, the complex coordinate transformation method was employed to add one angular momentum to the Schwarzschild–Tangherlini solution,  resulting in a rotating black hole.  Including a cosmological constant in Myers and Perry's solutions with arbitrary angular momenta in all higher dimensions, in both Kerr-Schild and Boyer-Lindquist form, was accomplished in \cite{Gibbons:2004js}. Note that the existence of a five-dimensional rotating black ring geometry \cite{Emparan:2001wn} shows that the Myers--Perry solution is not unique, unlike the four-dimensional Kerr solution. As expected from the four-dimensional case (where only the Kerr--Newman solution exists), adding physical sources to vacuum rotating black holes is extremely difficult. Going beyond Einstein theory, charged rotating  black holes have been discussed in the framework of certain supergravity theories and string theory, e.g.  \cite{YOUM19991, Gauntlett_1999, ChongPRD2005, ChongPRL2005, PhysRevD.95.064018}.

Since solving the Einstein equations directly in the fully rotating case especially in higher dimensions (even restricting to specific dimensions five, six or separating even and  odd dimensions) proves to be very challenging, one possible way around this problem is to generalize successful generating techniques, such as Newman--Janis trick \cite{Erbin:2014lwa, Erbin:2014aya, Erbin:2016lzq, Junior:2020lya, BMirza2020}. Apart from the original form of the Newman--Janis technique which involves a null tetrad basis transformation, there is another formulation of this algorithm proposed by Giampieri \cite{Giampieri1990}. It does not rely on the metric inversion nor on finding suitable null tetrad basis in order to apply a complex transformation, which is particularly helpful when working in an arbitrary number of dimensions. However, the resulting replacement
of the metric functions coincides with the original prescription.

Our aim here is to test the applicability of (generalized) Newman--Janis trick in higher dimensions when nonlinear electrodynamics (NLE) is considered as a source, similarly to the analysis carried out in four dimensions in \cite{Kubiznak:2022vft}, which showed that standard Newman--Janis trick fails for any theory other than Maxwell already in linear order in rotation, when only so-called restricted NLE theories (dependent on a single invariant) are considered.

Nonlinear electrodynamics (NLE) has a century long history and was introduced as a cure for the problem of a divergent field of a point charge and to provide satisfactory self-energy of charged particles. The best-known and frequently used form of the theory was introduced in 1934 by Born and Infeld \cite{Born:1934gh}, while Euler--Heisenberg electrodynamics (developed soon after \cite{Heisenberg:1936nmg}) incorporated quantum corrections. A concise overview of NLE, discussing its formulation and properties, was given in a book by Pleba\'{n}ski \cite{Plebanski:1976gy}. 

Coupling NLE to general relativity represents a challenging problem since now even the source is governed by nonlinear equations, therefore the majority of studies analyze highly symmetric and static situations. As we showed in a series of papers \cite{Tahamtan-NE:2016, Tahamtan:2020lvq, Kubiznak:2022vft, Tahamtan:2023tci}, if these symmetries are broken, majority of NLE models do not provide generalizations of many standard Einstein--Maxwell solutions or the systems of PDE cannot be integrated using methods known from Maxwell case. Since all the known NLE solutions in four dimensions are twist--free, it would be extremely valuable to obtain rotating generalizations of the static spherically symmetric cases thus providing an NLE version of the Kerr–Newman solution.  Although this task is quite formidable, there have been several attempts to find such solutions. For example in \cite{Diaz-fullRotating2022, Breton-fullRotating2024} the authors present a partial solution without having an explicit expression for the Lagrangian in terms of the electromagnetic invariants. The other approach to find rotating black hole solutions is using the Newman–-Janis trick \cite{lombardo2004newman}.  As we already mentioned earlier, this trick works perfectly for vacuum (Schwarzschild) and Maxwell charged (Reissner--Nordstr\"{o}m) solutions in four dimensions but this is not the case for NLE theories. Not to mention that many rotating NLE solutions obtained in the literature using this trick are not correct, the proof is shown in \cite{Kubiznak:2022vft} for four dimensions. Here we will show that using a higher-dimensional generalization of the Newman--Janis trick, specifically the so-called Giampieri formulation \cite{Giampieri1990, Erbin:2014aya}, one cannot generate rotating solutions for any restricted nontrivial NLE. Similarly to \cite{Kubiznak:2022vft} the slow-rotation approximation will be used in higher dimensions as well, considering equations up to linear order in the rotation parameter. 

Apart from potentially ruling out the above mentioned trick, the slow-rotation approximation also serves as a first step toward understanding the fully rotating case, as one would expect for any linear approximation of  a complex nonlinear theory. Extension to higher dimensions can shed light on the crucial points of the approach and also provide more planes of rotation to consider. In \cite{Kubiznak:2022vft} we found a unique NLE model (later called RegMax \cite{Hale:2023dpf, Hale:2024lzh}) for slowly rotating spacetime using a generalized Lense–Thirring ansatz in four dimensions.  We will use the generalization of this metric ansatz to higher dimensions, as previously used for the Einstein--Maxwell theory and in various higher curvature gravities, as well as in the presence of non-trivial matter \cite{Aliev2006, Gray:2021roq, Gray:2021toe}.

Note that the extension of the exceptional RegMax theory (specifically the form of the Lagrangian) beyond four dimensions is of course not unique \cite{Tahamtan3d:2024}. Therefore, searching for higher-dimensional slowly rotating solutions can help fix this non-uniqueness by selecting a Lagrangian admitting such solutions.

\section{Giampieri’s formulation} \label{JN_Algorithm}
A reformulation of the Newman--Janis (NJ) trick was proposed by Giampieri \cite{Giampieri1990}, which provides a simplification of this algorithm. This formulation is more practical for generalizing to higher dimensions \cite{Erbin:2014aya, Erbin:2014lwa}.  In this section, we review this approach and its limitations. We start by assuming a static spherically symmetric metric in four dimensions (and extend it to higher dimensions afterwards) and follow the steps outlined in \cite{Erbin:2014aya}. The line element is given by
\ba \label{metric=SSS}
ds^2&=&-f(r)\,dt^2+\frac{dr^2}{f(r)}+r^2\,d\Omega^2\ ,
\ea
where $f$ is the unknown metric function and $d\Omega^2=\left( d\theta^2+\sin^2{\theta}\,d\phi^2\right)$.  The first step is to write the above metric in the Eddington-Finkelstein form, 
\ba \label{nullCoor}
ds^2&=&-f(r)\,du^2-2\,du\,dr+r^2\,d\Omega^2\ ,
\ea
where $r$ is a null coordinate. The next step is to change the coordinates via complex transformation 
\ba\label{Imaginary-transf}
u=u'+i a \cos{\psi},\,\,r=r'-i a \cos{\psi}.
\ea
It is obvious that with the above choice of the coordinates the metric becomes complex, but using the following ansatz which is not a standard coordinate transformation
\be \label{adhoc}
i\,d \psi=\sin{\psi}\,d\phi\,,
\ee
with $\psi=\theta$, the metric becomes real
\ba \label{Corr}
ds^2&=&-\tilde{f}\,(du'-a \sin^2{\theta}\,d\phi )^2\nonumber\\
&&-2\,(du'-a \sin^2{\theta}\,d\phi )\,(dr'+a \sin^2{\theta}\,d\phi )+\rho^2\, d\Omega^2\,, \nonumber\\
\ea
in which $\tilde{f}$ can be a function of $r'$, $\theta$ and $\rho^2=r'^2+a^2\cos^2{\theta}$ {(note that this corresponds to the replacement $r^2 \rightarrow r\,r^*$ during complexification)}.  Finally, to arrive at the Boyer–Lindquist coordinates, the following changes are applied
\ba \label{Boyer–Lindquist}
du'=dt-\lambda(r')\,dr'\,,\quad\quad d\phi=d\phi'-\chi(r') \,dr'\,,
\ea
requiring that the only non-diagonal component of the metric is $g_{t\phi'}$. The resulting metric then takes the following form (dropping all the primes)
\ba \label{fullRotatingCorr}
ds^2&=&-\tilde{f}\,dt^2+\frac{\rho^2}{\Delta} dr^2+\rho^2\,d\theta^2+{\Sigma}\,\sin^2{\theta}\,d\phi^2\nonumber\\
&&+2a(\tilde{f}-1)\sin^2{\theta}\,dt\, d\phi
\ea
with
\ba
\Delta&=&\tilde{f}\rho^2+a^2\sin^2{\theta}\,,\label{Delta-4dim}\\
\Sigma&=&2a^2(\tilde{f}-1)\sin^2{\theta}+r^2+a^2.
\ea

The higher-dimensional form of this trick is described in \cite{Erbin:2014lwa}. The form of full rotating metric after all the transformations are applied for spherically symmetric metric in arbitrary dimension (for simplicity,  the authors in \cite{Erbin:2014lwa} consider the odd dimensions separately, but the final result can be extended also for even dimensions, see equation B.13 in \cite{Erbin:2014lwa}) is
\ba \label{fullRotatingCorrhigherD}
ds^2&=&-dt^2+(1-\tilde{f})\left( dt-\sum_{i=1}^m\,a_i\,\mu_{i}^2\,d\phi_i \right)^2+\frac{r^{2}\rho^2}{\Delta'} dr^2\nonumber\\
&&+\sum_{i=1}^m\left(r^2+a^2_i\right)\left( d\mu_{i}^2+\mu_{i}^2\,d\phi^2_i \right)^2\ ,
\ea
{where $m=\lfloor\frac{d-1}{2}\rfloor$ is the number of independent rotation parameters $a_i$ and 
\[\Delta'=\tilde{f}\rho^2+\prod_{i=1}^m\,(r^2+a_i^2)\sum_{i=1}^m\frac{a_i^2\,\mu_{i}^2}{r^2+a_i^2}\,.\]
Note that in four-dimensional spacetime the above expression for $\Delta'$ reduces to \eqref{Delta-4dim}.} If we consider terms up to first order in rotation parameters {in \eqref{fullRotatingCorrhigherD}} we obtain
\ba \label{SlowRotatingCorrhigherD}
ds^2&=&-f\,dt^2+2\,(f-1)\sum_{i} a_i\,\mu_{i}^2\,dt\,d\phi_i+\frac{dr^2}{f} \nonumber\\
&&+r^2\sum_{i}\left( d\mu_{i}^2+\mu_{i}^2\,d\phi^2_i \right)^2
\ea
where $f$ now corresponds to the static solution. 

For single rotation (see B.25 in \cite{Erbin:2014lwa}) one arrives at
\ba \label{fullRotatingCorrhigherD-onerotation}
ds^2&=&-\tilde{f}\,dt^2+\frac{r^{d-3}\rho^2}{\Delta} dr^2+\rho^2\,d\theta^2+{\Sigma}\,\sin^2{\theta}\,d\phi^2\nonumber\\
&&+2a(\tilde{f}-1)\sin^2{\theta}\,dt\, d\phi+r^2\cos^2{\theta}\, d\Omega_{d-4}^2\ ,
\ea
where $\Delta$ and $\Sigma$ have the same values as in four dimensions. In the above metric, if we consider terms up to first order in rotation parameter, the form of the metric coincides with \eqref{SlowRotatingCorrhigherD} when only one rotation parameter is assumed. Comparing these two metrics (\eqref{fullRotatingCorrhigherD} and \eqref{fullRotatingCorrhigherD-onerotation}) up to first order in the rotation parameter with the generalized Lense–Thirring metric \eqref{metric=Anz}, we see that $h=f-1$. As we will show in the Section \ref{section:slowly-rotating}, such restriction, following from the standard Giampieri’s formulation (NJ algorithm), is very strong and consistent only with the Maxwell theory.

In Giampieri’s formulation, the function $\tilde{f}$ appearing above needs to be complexified as many times as the number of independent planes of rotation. In dimensions higher than five this leads to problems compared to relatively straightforward five-dimensional case which leads to Myers–Perry solution. Finding the correct complexification seems very challenging and this worsens for NLE which typically produces complicated forms of static metric function $f$.

Generalization of \eqref{Imaginary-transf} by introducing additional arbitrary functions of $r$ was attempted for NLE in four dimensions, but comparison with results for slowly rotating solutions \cite{Kubiznak:2022vft} showed that any such generalization cannot yield valid solution even in the linear order of rotation parameter. In the following we will use slow rotation approximation to explicitly show the limits of the above described tricks for the higher-dimensional NLE black holes.

Above we only discussed the transformation of the geometry represented by the metric.  We will now briefly show how the above described transformation (Giampieri's trick) modifies the vector potential in four--dimensional spacetime. We start with the following electric vector potential ansatz for static spherically symmetric spacetime
\be 
A=U(r)\,dt\,.
\ee
Expressing this potential in the Eddington-Finkelstein coordinates and removing a pure gauge term (as done in \cite{Erbin:2014lwa}) one obtains $A=U(r)\,du$. Now, one applies the same complexification and transformation rules as done for the metric \eqref{Imaginary-transf}-\eqref{adhoc} and \eqref{Boyer–Lindquist}, to finally arrive at 
\be 
\tilde{A}=\tilde{U}\,\left(dt-a \sin^2{\theta}\,d\phi'-(\lambda-\chi\,a\sin^2{\theta})\,dr'\right)\,.
\ee
Note that for Maxwell theory the expression $(\lambda-\chi\,a\sin^2{\theta})$ appearing in  the last term is just $r'$ dependent, therefore, the $\tilde{A}_{r'}$ component can be removed by a gauge transformation. {The $r'$ dependence of $\tilde{A}_{\phi'}$ component is completely determined by the static potential $\tilde{U}$. In the slow rotation approximation (described later) we generalize this component by considering additional function $\omega(r')$, thus working with the four-potential 
\[\Hat{A}=\tilde{U}\,\left(dt-\omega\, a \sin^2{\theta}\,d\phi'\right)\,.\]
Therefore, in four dimensions the above rotation-inducing transformation automatically leads to $\omega=1$. As we will show below, this is automatically the case in higher dimensions as well, thus severely restricting possible NLE models compatible with such prescription.} 


\section{Theories of nonlinear electrodynamics}
\label{sec2}
Theories of nonlinear electrodynamics that are minimally coupled to Einstein's gravity are described by the following action:
\begin{equation}\label{bulkAct}
    I= \frac{1}{16\pi} \int_{M} d^dx \sqrt{-g}\left(R +4{\cal L}-2\Lambda\right)\,,
\end{equation}
including a possibility for nontrivial cosmological constant $\Lambda$, which we parametrize as 
\be
\Lambda=-\frac{(d-1)(d-2)}{2\,\ell^2}\,, 
\ee
in terms of AdS radius $\ell$ when $\Lambda<0$. Here, ${\cal L}$ is the electromagnetic Lagrangian, which is taken to be a function of electromagnetic invariants (whose number depends on the dimension) with the crucial one being
\be
{\cal S}=\frac{1}{2}F_{\mu\nu}F^{\mu\nu}\,,
\ee
where, as always, we have $F_{\mu\nu}=\partial_\mu A_\nu-\partial_\nu A_\mu$, in terms of the vector potential $A_\mu$.  Only the {\em restricted class} of NLE theories, obtained by considering solely the invariant ${\cal S}$ \footnote{Note that the number of independent invariants of electromagnetic field grows as $[d/2]$ in higher dimensions.}: 
\be\label{restrict}
{\cal L}={\cal L}({\cal S})\,, 
\ee
is considered in this paper.  Introducing the following notation:
\be
{\cal L}_{\cal S}=\frac{\partial {\cal L}}{\partial {\cal S}}\,,
\ee
the {\em generalized Maxwell} equations read
\be\label{FE}
d*D=0\,,\quad   dF=0\,, 
\ee
where 
\be\label{Edef}
D_{\mu\nu} = -2\,\frac{\partial \mathcal{L}}{\partial F^{\mu\nu}}
=-2\,\Bigl({\cal L_S}F_{\mu\nu}\Bigr)\,. 
\ee
We also obtain the following 
{\em Einstein equations}: 
\be \label{Hmunu}
G_{\mu\nu}+\Lambda g_{\mu\nu}=8\pi T_{\mu\nu}\,,
\ee
where the generalized electromagnetic energy-momentum tensor reads 
\be\label{Tmunu}
T^{\mu\nu}=-\frac{1}{4\pi}\Bigl(2F^{\mu\sigma}F^{\nu}{}_\sigma {\cal L_S}-{\cal L}g^{\mu\nu}\Bigr)\,.
\ee

The corresponding equations of motion straightforwardly follow from the above.

\section{ SLOWLY ROTATING NLE SOLUTIONS}\label{section:slowly-rotating}
In this and the next sections, we show how one can construct slowly rotating spacetimes coupled to any NLE for arbitrary dimensions. {We will only consider electrically  charged solutions since the magnetic monopoles in higher dimensions represent quite a nontrivial problem due to topological obstructions, these can be solved in various ways (e.g., using potential of higher rank or extended objects \cite{Nepomechie:1984wu}) and would require reformulation of the description of the electromagnetic field which goes beyond this study.}

{The well known metric used to solve slowly rotating problem in four dimensions is  the Lense–Thirring ansatz \cite{lense1918influence}, which is a slow rotation approximation of the Kerr metric \cite{Kerr1963}. The Lense–Thirring metric was generalized to higher dimensions and frequently used in the literature, e.g. \cite{Gray:2021roq}, \cite{Gray:2021toe}.} The generalized Lense–Thirring ansatz for arbitrary d-dimensions is given by
\ba \label{metric=Anz}
ds^2&=&-f(r)\,dt^2+\frac{dr^2}{f(r)}+r^2 \sum^{m}_{i=1}  \left( d\mu_{i}^2+{\mu_{i}^2}\,d\phi^2_i \right)+\epsilon\,r^2\, d\nu^2  \nonumber \\
&&+\sum^{m}_{i=1} a_i\mu_{i}^2  h(r)dt d\phi_i\,,
\ea
for $m=\lfloor\frac{d-1}{2}\rfloor$  independent rotation parameters $a_i$. The functions $f, h$  are functions of the radial coordinate $r$, and the coordinates $\mu_i$ and  $\nu$ obey
\[\sum^{m}_{i=1}\mu_{i}^2+\epsilon\, \nu^2=1\]
where $\epsilon= 1, 0$ in even and odd dimensions, respectively. Note that $\epsilon=1$ case can be incorporated into \eqref{SlowRotatingCorrhigherD} by considering $\mu_{m+1}=\nu$ and $\phi_{m+1}=0$. 

The corresponding electromagnetic field potential in $d$ dimensions has the form
\be \label{A}
A=U(r)\left(dt-\sum^{m}_{i=1}a_i\,\mu_{i}^2 \,\omega(r) d\phi_i \right)\,,
\ee	
 where $U$ corresponds to the static solution and $\omega = \omega(r)$ captures the effect of rotation. The standard electromagnetic scalar invariant corresponding to the above electromagnetic potential  is given by
\be\label{inv}
{\cal S}=-(U_{,r})^2\,+O(a_i^2)\,.
\ee

For an arbitrary Lagrangian ${\cal L({\cal S})}$, the two independent diagonal components {(denoted by $H_{\mu\nu}$)} of the Einstein equations \eqref{Hmunu} up to the first order in rotation parameters are the following 
\ba
&&H_{rr}=8\,r^2\,{\cal L}_{\cal S} \left( U_{,r} \right)^2+2\,r^2(2{\cal L}-\Lambda)\\
&&-r(d-2)\,f_{,r}-\left( d-2 \right)\left( d-3 \right)\Big(f-1\Big)=0 \nonumber
\ea
and 
\ba
&&H_{\theta \theta}=2\,r^2(2{\cal L}-\Lambda)\\
&&-r^2\,f_{,rr}-2r(d-3)\,f_{,r}-\left( d-4 \right)\left( d-3 \right)\Big(f-1\Big)=0\,.\nonumber
\ea
 Combining the two equations above, we get
 \ba \label{finding-f}
&& 8\,r^2{\cal L}_{\cal S} \left( U_{,r} \right)^2\\
&& +{r}^{2}f_{,rr}+ r\left( d-4\right)f_{,r}-2\, \left( d-3 \right) \Big(f-1\Big)=0 \nonumber
\ea
where we can find the static metric function $f$ in terms of the static potential $U$. The modified Maxwell equation \eqref{FE} in $d$-dimensions
\be \label{LF(r)}
(\nabla \cdot D)_t=\Bigl(r^{(d-2)}\,U_{,r}{\cal L}_{\cal S}\Bigr)_{,r}=0\,
\ee  
leads to the following result
\be\label{LSformula}
{\cal L}_{\cal S}=\frac{\beta}{r^{(d-2)}\,U_{,r}}\,
\ee
where $\beta$ is an integration constant.  
Combined with \eqref{inv} (and upon fixing the Lagrangian ${\cal L}$) this yields the equation for the potential $U(r)$. However, we can insert ${\cal L}_{\cal S}$ from \eqref{LSformula} into \eqref{finding-f} and solve the differential equation to obtain  
\ba\label{static-f}
 f=1+f_2\,r^2+\frac{f_1}{r^{d-3}}+\frac{8\beta\,U}{(d-1)r^{d-3}}-\frac{8\beta\,r^2}{d-1}\int \frac{U_{,r}dr}{r^{d-1}} \,.\nonumber\\
 \ea
 Here $f_1$ and $f_2$ are integration constants related to the mass parameter and the cosmological constant. Note that these functions, $f$ and $U$, correspond to the static solution.  
 
 From the remaining Einstein equations \eqref{Hmunu}, we will find the metric function $h$ \eqref{metric=Anz} and the function $\omega$ in the vector potential \eqref{A}. Naturally, since these two functions encode the effect of rotation at linear order, the off--diagonal terms of \eqref{Hmunu} are used. From  $H_{t\phi}$ we find 
\ba \label{Htphi}
8r^2{\cal L}_{\cal S} U_{,r} \left( U\,\omega\right)_ {,r}+{r}^{2}h_{,rr}+ r\left( d-4
\right)h_{,r}-2\left( d-3 \right)h=0\nonumber\\
\ea
that represents a differential equation for the metric function $h$ with the source term given by the electric potential $U$ and $\omega$. This equation can be integrated to give
 \ba \label{h-with-omega}
 h=h_2\,r^2+\frac{h_1}{r^{d-3}}+\frac{8\beta\,U\,\omega}{(d-1)r^{d-3}}-\frac{8\beta\,r^2}{d-1}\int \frac{\left( U\,\omega\right)_ {,r}}{r^{d-1}} dr\, ,\nonumber\\
 \ea
where $h_1$ and $h_2$ are integration constants. One should set $h_2=0$ since its non-trivial value induces rotation at infinity which we want to avoid. One simple case in which equation \eqref{h-with-omega} can be integrated is when $U\omega=const.$  but such a choice {(from NLE equation \eqref{FE})} leads to $U \approx r^{d-3}$ which is not physically interesting. 

The last remaining equation, which comes from another component of NLE equation \eqref{FE}, can be used to find $\omega$ in terms of $U$. Note that if $\omega$ would be constant, this equation would serve as a constraint instead:
\ba\label{NLE-general-Omega}
&&r^2fU\left(\omega_{,rr}U_{,r}-\omega_{,r}U_{,rr}\right)+ r^2\,U^2_{,r}\left(2f\,\omega_{,r}+{r^2}\omega\Bigl(\frac{f}{r^2}\Bigr)_{,r}\right)\nonumber \\
&&-\left(2(d-3)-r^4\omega_{,r}\Bigl(\frac{f}{r^2}\Bigr)_{,r}\right) UU_{,r}+rU^2_{,r}\left(2h-rh_{,r}\right)=0\,.\nonumber\\
\ea

In order to obtain all the unknown functions, both static and rotating, one needs to specify the electromagnetic Lagrangian or at least the electric potential. It is clear that the Maxwell potential has the simplest form (in terms of $r$), and even with this simple Maxwell potential one cannot find the general expression for the function $\omega(r)$ except for four dimensions. The special case in which $\omega(r)=const.$ leads to either the Maxwell theory or a special form of NLE, as we will see in the following sections.

\subsection{Maxwell uniqueness in all dimensions} \label{Maxwell-uniqueness}

In our previous results for four dimensions \cite{Kubiznak:2022vft}, we proved a  “No Go theorem” that, in particular, rules out the standard Newman–Janis trick as a way of deriving rotating NLE solutions within a large family of NLE models. We can extend this theorem for arbitrary dimension ($d \geq 4$). Note that as we have already shown in the Section \ref{JN_Algorithm},  for arbitrary dimension the relation $h=f-1$  holds at the linear level of rotations.  

Starting with the condition $h=f-1$, and inserting the ${\cal L}_{\cal S}$  from \eqref{finding-f} to \eqref{Htphi}, finally we get 
\ba\label{Maxwell-uniq}
[U(\omega-1)]_{,r}\,\zeta=0
\ea
where $\zeta=(r^2\,f_{,rr}+r(d-4)\,f_{,r}-2(d-3)(f-1))/U_{,r}$.  The above equation \eqref{Maxwell-uniq} is basically the same as in four dimensions, see the equation (48) in \cite{Kubiznak:2022vft}. This is because the role of the extra dimensions appears only in $\zeta$, which does not affect our final results. Therefore, with the same arguments as in four dimensions\footnote{Note that the other solution for equation \eqref{Maxwell-uniq} is when $\zeta=0$. With this choice, the static metric solution would be only vacuum one, namely $f=1-\frac{f_0}{r^{d-3}}+\Lambda\,r^2$. Basically, this solution is not allowing any sources.}, we conclude that 
\[\omega=1,\qquad U=\frac{q}{r^{d-3}}.\]

Therefore, our theorem extends as following:

\begin{theorem}[Maxwell uniqueness]{For the restricted class of theories, ${\cal L}({\cal S})$, in arbitrary dimensions the only NLE consistent with $h=f-1$ for the ansatz \eqref{metric=Anz} and \eqref{A} (and thence consistent with the generalized NJ trick) is the Maxwell theory.}
\end{theorem}

The result and the theorem are the same as in four dimensions;  therefore, we can extend the failure of Newman–Janis trick for generating rotating electrically charged spacetimes in $d \geq 4$ from NLE models at the linear level. {Note that although the theorem above singles out Maxwell theory as the only one compatible with the slow rotation approximation consistent with NJ trick, it does not mean that this remains true beyond the linear order in the rotation parameters.}

\subsection{Maxwell theory and  $h \neq f-1$ } \label{Max-omega(r)}

In the previous part, first we imposed the condition $h=f-1$ and then we found the uniqueness of  Maxwell theory.  Here, we assume the $h \neq f-1$ condition and investigate whether the function $\omega$ is constant or not for the Maxwell theory. The purpose of this section is to highlight that this process is not straightforward even in this case before moving on to NLE. 

Since the static electric potential is known in Maxwell theory, the only remaining unknown function to be found is $\omega$.  The static electric Maxwell potential for arbitrary dimensions $d \geq 4$ is 
\[U=\frac{{q'}}{r^{d-3}}\,.\]
Note that for Maxwell theory the ${\cal L}_{\cal S}={-1/2}$, where from \eqref{LSformula}, we see that $\beta={q'}\,(d-3){/2}$ {(where $q'=\sqrt{2}\,q$)}.  By inserting the above potential in the equation \eqref{NLE-general-Omega} and assuming $f_2=0$ for simplicity (thus ignoring cosmological constant), we get
\begin{eqnarray}\label{Max-omega}
&r^2\omega_{,rr}\bigl[ 4q^2\,(d-3)R^{-1} +(d-2)\left(f_1+R\right) \bigr]\\ \nonumber
&-r\omega_{,r}\bigl[4q^2\,(d-3)(3d-8)R^{-1}+\bigr. \\ \nonumber
&\bigl.(d-2)\bigl((2d-5)f_1+(d-2)R\bigr)\bigr]\\ \nonumber
&+(d-3)(d-2)(d-1)\bigl(\omega\,f_1-h_1\bigr)=0
\end{eqnarray}
where $R \equiv r^{d-3}$.  Since the above equation is not solvable analytically, we use an approximation for large $r$ or in other words an asymptotic expansion in $r$. In this approximation method we assumed the solution in form of a power series (using first 5 orders). When only the dominant term of equation \eqref{Max-omega} is assumed to vanish the resulting series for $\omega$ around infinity becomes
\ba \label{series-omega-Maxwell}
\omega=1-\frac{(d-1)}{2\,(d-2)} \frac{(f_1-h_1)}{r^{d-3}}+\omega_1\,\frac{f_1(f_1-h_1)}{r^{2d-6}}+\cdots\nonumber\\
\ea
where $\omega_1$ is a constant. Demanding the equation to be valid in next two subleading orders we obtain $f_1=h_1$, unless special relation between $q$ and $f_{1}$ is imposed (which would essentially result in mass being fixed by charge) covering only single parameter family of solutions. Note that considering $f_1=h_1$ means $h=f-1$ and as shown in \eqref{Maxwell-uniq} we necessarily have $\omega=1$. Assuming $\omega=1$ in the equation \eqref{Max-omega} one immediately obtains $f_1=h_1$, which means $h=f-1$. 

One should notice that the above results are obtained as an approximation for large $r$ since the equation \eqref{Max-omega} is not solvable for arbitrary dimension $d$. The exception is for four dimensions, where one is able to find analytical solution (note that we are studying the Maxwell theory). Indeed, when $d=4$, the equation \eqref{Max-omega} reduces to
{\ba
&&r\left(2q^2+f_1r+r^2\right)\omega_{,rr}-\left(8q^2+3f_1r+2r^2\right)\omega_{,r} \nonumber\\
&&+3\left(\omega\,f_1-h_1\right)=0 
\ea}
and solving it for $\omega$ one obtains
\ba
\omega&=&\frac{h_1}{f_1}+\omega_1\tilde{\omega}+\omega_2\left[12f_1\tilde{\omega} \arctan{\frac{f_1+2r}{\sqrt{8q^2-f^2_1}}} \nonumber \right.\\
&&\left. +\left(128q^4+8q^2f^2_1+3f^3_1r-6f^2_1r^2\right)\sqrt{8q^2-f^2_1}\right]. \nonumber
\ea
where $\tilde{\omega}=\left(16q^4+6q^2f_1r-f_1r^3\right)$ and the two coefficients $\omega_1$ and $\omega_2$ are integration constants. In order to have reasonable asymptotic behavior for $\omega$, these two constants should be zero. Therefore, we see that although we start with $\omega \neq \text{constant}$, we find analytically that the only solution for the Maxwell theory has $\omega=\text{constant}$. This agrees with the asymptotic method found in \eqref{series-omega-Maxwell}.

\subsection{Special NLE models with $\omega(r)=1$ }

As we already showed in the previous parts, if $h=f-1$ holds we obtain $\omega=1$ and Maxwell theory is the only one that satisfies the field equations. Now the question is: if we assume only $\omega=1$ (while $h \neq f-1$), is there any nontrivial NLE?

The electromagnetic potential \eqref{A}, considering $\omega=1$, is now 
\be \label{A-omega=1}
A=U(r)\left(dt-\sum^{m}_{i=1}a_i\,\mu_{i}^2 \,d\phi_i \right)\,.
\ee	
To illustrate the explicit form of the electromagnetic potential we show the electromagnetic potential for four, five and six dimensions. The electromagnetic potential in four dimensions is
\[ A=U(r)\left(dt-a\, \sin^2{\theta} \,d\phi\right),\]
for five dimensions
\[ A=U(r)\left(dt-a\, \sin^2{\theta} \,d\phi-b\, \cos^2{\theta} \,d\psi \right),\]
and finally for six dimensions
\[ A=U(r)\left(dt-\sin^2{\psi}\,(a\, \sin^2{\theta}\,d\phi+b\, \cos^2{\theta} \,d\psi ) \right).\]
Since we express all the equations in terms of $U$ (see \eqref{static-f}, \eqref{h-with-omega}, \eqref{NLE-general-Omega}, while imposing $\omega=1$) the only remaining unknown  function is the static electric potential $U$. Setting $\omega=1$ in \eqref{h-with-omega} gives
\ba
 h=h_2\,r^2+\frac{h_1}{r^{d-3}}+\frac{{8}\beta\,U}{(d-1)r^{d-3}}-\frac{{8}\beta\,r^2}{d-1}\int \frac{U_{,r}dr}{r^{d-1}} \nonumber\\
 \ea
and we can see the following relation between $f$ and $h$ 
\begin{equation}\label{fh}
f-h=1+r^2\Big(f_2-h_2\Big)+\frac{f_1-h_1}{r^{d-3}}\,.
\end{equation}
If we plug all these expressions into the second NLE equation \eqref{FE}, or equivalently \eqref{NLE-general-Omega} with $\omega=1$ inserted, we obtain the following result

\ba \label{NLE-2}
\frac{U_{,r}}{U} =\frac{2( d-3)}{r^4\Bigl(\frac{f-h}{r^2}\Bigr)_{,r}}\,.
\ea
Knowing the relation between $f$ and $h$, namely the equation \eqref{fh} and the above equation \eqref{NLE-2}, we realize that the only electric potentials admitting the slowly rotating solutions are

\begin{equation} \label{generalPotential}
	U(r)={\frac {{Q}}{{r}^{d-3}+ \frac{\left( d-1 \right)\left( f_1-h_1 \right)  }{2}}}
\end{equation}
where $Q$ is an integration constant. For simplicity, we consider $f_2=h_2=0$, disregarding cosmological constant again. The other two constants, $f_1$ and $h_1$, can have different values, which leads to different results; for example, if we assume $f_1=h_1$ (which means $h=f-1$), then the above potential is the Maxwell potential for arbitrary dimensions, $d \geq 4$,  which agrees with our extended theorem for $d \geq 4$ given in Section \ref{Maxwell-uniqueness}.  If we consider the case when  $f_1 \neq h_1$,  then this choice leads to a specific NLE theory. In four dimensions, with $f_1 \neq h_1$ and choosing  the constants  $Q=1$ and $M_0=\frac{f_1-h_1}{2}$, the electric potential \eqref{generalPotential} coincides with the potential (56) in \cite{Kubiznak:2022vft}. The corresponding NLE Lagrangian that gives this electric potential is called RegMax. In other words, in four dimensions, there are two theories that yield the Lense–Thirring solutions with $\omega=1$: Maxwell theory and RegMax NLE.

For higher dimensions it is not easy to express the NLE Lagrangian explicitly in terms of the electromagnetic invariant ${\cal S}$.  One potential candidate would be RegMax in higher dimensions, but generalization of RegMax to other dimensions is not unique as pointed out in \cite{Tahamtan3d:2024}. One of the possible generalizations corresponds to \eqref{generalPotential}; however, in this way it is hard to find $r$ in terms of  ${\cal S}$ (since one obtains high-order polynomial equation), and hence to find the Lagrangian in terms of the invariant ${\cal S}$. Another way to generalize RegMax to higher dimensions is to use the electric field of the form $E(r)=\frac{Q}{(r+r_0)^{(d-2)}}$; a procedure to find the Lagrangian for arbitrary dimensions is shown in \cite{Tahamtan3d:2024}. This generalization clearly does not fit with the potential \eqref{generalPotential}. Therefore, slowly rotating case with $\omega=1$ can be used to select preferred higher-dimensional generalization of RegMax NLE replicating one of the possible ways in which four-dimensional RegMax was discovered.

Although finding the Lagrangian in terms of electromagnetic invariant ${\cal S}$ for $d>4$ is difficult, it is trivial to find it in terms of $r$. First, we denote $r_0= \frac{\left( d-1 \right)\left( f_1-h_1 \right)}{2}$; then, knowing the relation between the electric potential and ${\cal L}_{\cal S}$ from \eqref{LSformula}, we find the  following expression for the Lagrangian
\ba\label{Lagrangian}
&&{\cal L}(r)=\frac{2Q\beta}{r_0} \left[ (d-1)(d-2)\int {\frac{dr}{r^d+r_0r^3}} \right.\nonumber\\
&&\left. +\frac{r\left(r^d(d-2)+r_0r^3(2d-5)\right)}{(r^d+r_0r^3)^2}\right] \,.
\ea
Note that it is possible to adjust the constants $\beta$ and $r_0$ in order to have Maxwell limit and obtain a parameter analogous to the Born–Infeld parameter, known as the maximal field strength allowed in the theory. 

Nevertheless, here we are interested in finding slowly rotating solutions corresponding to any NLE no matter what is the form of Lagrangian.  It seems that the only existing slowly rotating solutions for any NLE theory (limited to ${\cal L}({\cal S})$ theory and $\omega=1$) are those corresponding to above electromagnetic potential given in \eqref{generalPotential}.

\subsection{ Asymptotic behavior and regularity of $\omega(r)$ }
In order to find some NLE models corresponding to general $\omega(r)$, we need to solve equation \eqref{NLE-general-Omega}.  The general expressions for metric functions $f$ and $h$ in terms of $\omega$ and $U$ are given in \eqref{static-f} and \eqref{h-with-omega} but solving the ODE \eqref{NLE-general-Omega} is not easy.  If we assume that the mass-like terms in $f$ and $h$ vanish, meaning $f_1=0$ and $h_1=0$, the equation simplifies substantially. We have tried for several specific NLE models with these assumptions, but we were able to obtain exact solutions only for the conformal Maxwell theory (${\cal L}\sim {\cal S}^{d/4}$).  Unfortunately these kinds of solutions, with the condition $f_1=0$ and $h_1=0$, do not represent a black hole solutions, therefore we are not interested in them.

Since finding $\omega(r)$ exactly is impossible, one approach is to solve the equation \eqref{NLE-general-Omega} asymptotically as we already discussed for Maxwell theory in Section \ref{Max-omega(r)}. We use a series expansion for $\omega$ in two regions: small $r$, expansion around zero, and large $r$, asymptotic expansion.  We start by assuming the form of the electromagnetic potential corresponding to the region considered (around zero or at infinity), and then find the series coefficients of $\omega$. To make the equations shorter we also assume $f_2=h_2=0$ in \eqref{static-f} and \eqref{h-with-omega}.

\subsubsection{solution for $\omega$ near infinity}
Since we are interested in NLE models which have a Maxwell limit such as Born--Infeld or RegMax model, etc., we assume the electromagnetic potential in the following form (capturing the first two higher order corrections)
\ba \label{U-App-Inf}
U=\frac{q_1}{r^{d-3}}+\frac{q_2}{r^{d-2}}+\frac{q_3}{r^{d-1}}\ ,
\ea
by using \eqref{NLE-general-Omega} and considering only higher order terms in $r$, we find the approximate expression for $\omega$ for $d\geq 4$
\ba \label{omega-App-Inf}
\omega=\omega_0+\frac{\omega_1}{r}+\frac{\omega_2}{r^{2}}+\frac{\omega_3}{r^{3}} + \cdots
\ea
in which $\omega_i$ coefficients are determined by demanding \eqref{NLE-general-Omega} to be satisfied in the highest orders.  If we start with an electromagnetic potential without a Maxwell limit (such as the conformal Maxwell model in arbitrary dimension $d>4$, or regular conformal Maxwell one, \cite{Tahamtan3d:2024}), then the $\omega_0$ term vanishes. 

Let us consider explicit dimensions in \eqref{U-App-Inf}; for example, for $d=4$ dimensions,  the coefficients for $\omega$ in \eqref{omega-App-Inf} up to second order are (with $\omega_0=1$)
\ba
 \omega_1&=&-\frac{3}{4}\left(f_1-h_1\right)-\frac{q_2}{2q_1}\,,\nonumber\\
\omega_2&=&\frac{3\,f_1}{5}\left(f_1-h_1\right)+\frac{q_2}{20q_1}\left(5f_1+3h_1\right)-\frac{4q_1q_3-3q^2_2}{10q^2_1}\,.\nonumber\\
\ea
It is clear that when $q_2=0$ and $q_3=0$ (i.e., when the electric potential is just Maxwell) and $f_1=h_1$ is satisfied, the function $\omega$ is constant, consistent with previous discussion after equation \eqref{series-omega-Maxwell}.  For dimensions  other than four, the procedure is the same.  For example, if $d=5$, these coefficients up to third order are  (with $\omega_0=1$)
\ba
\omega_1&=&-\frac{2q_2}{5q_1}\,,\,\,\,\omega_2=-\frac{2}{3}\left(f_1-h_1\right)-\frac{4q_1q_3-3q^2_2}{12q^2_1}, \nonumber\\
\omega_3&=&\frac{2q_2}{21q_1}\left(2f_1+h_1\right)+\frac{q_2\left(172q_1q_3-69q^2_2\right)}{420q^3_1}.
\ea
Similarly to these two examples for four and five dimensions, it is also possible to find the coefficients $\omega_i$ for other dimensions.

\subsubsection{solution for $\omega$ around the origin}
In this part we use the approximation around the origin. Here we assume NLE models that regularize the point charge field at the origin (as was the original motivation for developing NLE theory) and therefore assume polynomial behavior of the following form
\ba
U=u_0+u_1\,r+u_2\,r^2+ \cdots\,.
\ea

In fact, we consider stronger regularity at zero by removing the linear term ($u_{1}=0$) to ensure the regularity of the electric field. Inserting the above electromagnetic potential into the equation \eqref{NLE-general-Omega}  (and considering only highest order terms), the $\omega$ function takes the form
\ba
\omega=\omega_0+\omega_1\,r+\omega_2\,r^2+ \cdots
\ea
As in the approximation used around infinity in the previous section, the form of $\omega$ and the potential $U$ are similar.  One can find explicit expressions for the series coefficients $\omega_i$ by again demanding that \eqref{NLE-general-Omega} be satisfied at several leading orders.

Since we are not able to solve equation \eqref{NLE-general-Omega} and find the exact form of $\omega(r)$, we used an approximation method.  Therefore, we can not state anything about the uniqueness of  $\omega =\text{const.}$ for NLE models.

\section{Summary}
In this paper, we have presented a generalized Lense–Thirring ansatz \eqref{metric=Anz}, together with the corresponding electromagnetic vector potential, \eqref{A}. We were able to find (at least in principle) the
corresponding Einstein--NLE equations to linear order in the rotation parameter. 

Moreover, we have extended the “No Go Theorem”  (the original one was found for four dimensions in \cite{Kubiznak:2022vft} ) which, similarly to the four-dimensional case, establishes that the Maxwell theory is the only NLE among all theories of the form ${\cal{L}}({\cal{S}})$ that admits a function $h$ given by the “natural” expression $h=f-1$. This shows that the standard rotation generating method -- such as Giampieri's formulation (a method well suited to higher dimensions) -- while successful in obtaining the fully rotating vacuum and Maxwell solutions (although only slowly rotating or Yang--Mills charged cases are known in higher dimensions), fails to produce fully rotating solutions for any nontrivial NLE in four and higher dimensions. This statement was proven by comparing the results at the level of the slow rotation approximation. One future task is to modify this type of generating methods in order to produce the correct fully rotating NLE solutions and derive exact Maxwell-charged rotating black hole in higher dimensions.

We have shown that when $\omega=1$, there are special NLE models in all dimensions providing slowly rotating solutions. We know that, in four dimensions, this NLE model is RegMax, but in higher dimensions we could not express the Lagrangian in terms of the electromagnetic invariant explicitly due to the involvement of high-order polynomials. The compact form of the Lagrangian for all dimensions in terms of $r$ is given in \eqref{Lagrangian}.

In the last part of the paper, we tried to analyze slowly rotating solutions when $\omega$ is not a constant. Since finding exact solutions, even with the imposition of the Lense-Thirring ansatz for the metric, was not possible, we used an approximation method. More specifically, we have chosen the form of the static electric potential to determine the form of $\omega$ near infinity and around the origin when regularity was assumed.

We hope that the obtained results for slowly rotating solutions with NLE sources will provide a hint for finding the fully rotating ones. At the same time, they should serve as a `warning' regarding the standard rotation generating trick.

\section*{Acknowledgment}
The author would like to thank David Kubiz\v n\'ak and Otakar Sv{\'i}tek for helpful discussions. {We are
as well grateful to the referee for numerous improvements of this manuscript.}

\bibliography{references}
\bibliographystyle{JHEP}

\end{document}